\documentstyle{caosp}
\begin{document}
\pubyear{1998}
\volume{27}
\firstpage{332}
\hauthor{I.V. Chernyshova {\it et al.}}
\title{The unusual A-star VW Ari: chemical composition revisited}
\author{I.V. Chernyshova \and S.M. Andrievsky
\and V.V. Kovtyukh \and D.E. Mkrtichian }
\institute{Department of Astronomy, Odessa State University, Shevchenko Park,
270014, Odessa, Ukraine}
\maketitle
\begin{abstract}
Seven high-resolution and high S/N CCD spectra were used to derive elemental
abundances in the atmosphere of VW Ari A (T$_{eff}$=7200, $log~g$=3.7) which is
the primary component of a visual binary system. The synthetic spectrum
technique applied in the analysis allowed to reveal the following feature:
the atmosphere
of this star is strongly deficient in some metals, while light elements have
solar-like abundances. Taking into account these results, one can suggest
that VW Ari A is a $\lambda$ Boo-type star. Another argument supporting this
supposition is the following: on the diagrams ``$(b-y)-c_{1}$'', 
``$(b-y)-m_{1}$'' and ``$(b-y)-\beta$'' (Paunzen et al. 1997) VW Ari A falls 
exactly in the region occupied by the $\lambda$ Boo stars.

Note also, that a previous analysis (Andrievsky et al. 1995) has shown that
the secondary component of VW Ari has a normal metallicity. Differences
in chemical compositions of the two components appear to be due to the specific
evolution of the primary VW Ari A.
\keywords{Stars: individual: VW Ari -- Stars: chemically peculiar}
\end{abstract}
\section{Introduction}
VW Ari (HD 15165, BDS 1269) is a remarkable visual binary system consisting of
two components: VW Ari A (V=6.$^{m}$71, A-type) and its companion
(V=8.$^{m}$33, F-type). The primary VW Ari A is a multiperiodic
pulsating star (probably of $\delta$ Sct-type) having non-radial modes.
This star shows the spectrum
typical of very metal-deficient stars. The rather high $v\sin i$ value
found for this star, makes it difficult to derive accurate elemental
abundances.
A first attempt was undertaken by Andrievsky et al. (1995), who showed that
calcium and iron are strongly deficient in the atmosphere of VW Ari A, while
the secondary component possesses a solar-like chemical composition. Such a
strange discrepancy between the metallicities of the two components
can be explained by several
hypotheses. For example, these stars possibly do not constitute a physical pair
or, in case they do, such an unusual stellar system could be formed as a result
of stellar capture.

Nevertheless, taking into account that 1) with a rather high probability VW Ari
is a binary system and 2) the probability of stellar capture in the field
is too small, we propose that the difference in chemical composition of both
components could appear simply due to the peculiar evolution of VW Ari A as a
$\lambda$ Boo-type star.

The atmospheres of this type of stars are known to be strongly deficient in
some heavy metals, while CNO-elements exhibit solar-like abundances (see
e.g. St\"urenburg, 1993).

To check this hypothesis, we performed a detailed spectroscopic analysis of VW
Ari
(primary component of the system) based on the spectral synthesis technique.
\section{Observation}
Seven CCD spectra have been obtained on 21 November 1994 with the \'echelle
spectrometer LYNX (modified version: 29 spectral orders with the length of
each order $\approx$ 60 \AA) on the 6-m telescope (Special Astrophysical
Observatory of the Russian Academy of Sciences, Russia, Northern Caucasus).
The detailed description of the spectrometer is given by Panchuk et al. (1993).
The resolving power was 24000, S/N $\approx$ 100.
The spectral region was 5035-7185 \AA.
The epochs at mid-exposures were the following: JD 2449670+ 1) 8.158,
2) 8.165, 3) 8.186, 4) 8.215, 5) 8.232, 6) 8.247, 7) 8.263.

All spectra have been reduced using the DECH20 code (Galazutdinov, 1992),
which includes extraction of spectra from images, dark and cosmic hits
subtraction, flat-field correction, wavelength calibration, etc.

\section{Atmospheric parameters}
The effective temperature and gravity for VW Ari A (T$_{eff}$=7200 K,
$log~g$=3.7) were estimated using the photometric indices $(b-y)=0.192$ and
$c_{1}=0.801$, and the calibration by Kurucz (1991). We adopted a microturbulent
velocity of 3 km\,s$^{-1}$, which is appropriate for A-F main-sequence stars, and
$v\sin i = 90$~km\,s$^{-1}$ was taken from Abt (1980).
\section{Method and results of the analysis}
The STARSP code (Tsymbal, 1996) was applied to derive the elemental abundances.
The atmosphere model was interpolated from Kurucz's (1992) grid. The
input oscillator strengths of the investigated lines and blends were initially
corrected by comparison of the solar synthetic spectrum (solar model from
Kurucz's grid, $V_{\rm t}=1$~km\,s$^{-1}$ and solar abundances from Grevesse and
Noels, 1993) with the solar flux spectrum (Kurucz et al. 1984).
The resulting abundances were found by means of the optimal fitting of the
synthetic spectrum to the observed one. They are given in Table 1.

\begin{table}[t]
\small
\begin{center}
\caption{Abundances for VW Ari A}
\label{t1}
\begin{tabular}{lccccccccccccc}
\hline
El.       &C  &  O &Na &Mg  &Si  & S &Ca  &Sc  &Ti   &Cr  &Fe  &Ni  &Ba  \\
$[$El/H$]$&0.0&-0.3&0.0&-1.5&-0.9&0.0&-1.0&-1.4&-0.50&-0.9&-1.6&-1.0&-0.8\\
\hline
\end{tabular}
\end{center}
\end{table}
\section{Discussion}
The abundance pattern in the atmosphere of VW Ari resembles that of $\lambda$
Boo-type stars (see, e.g. St\"urenburg, 1993, Andrievsky et al., 1998):
normal abundances (or slight underabundances) of carbon and oxygen and strong
deficiency of other elements.

An additional confirmation that VW Ari could be a $\lambda$ Boo star
is its position in $(b-y)-m_{1}-c_{1}-\beta$ diagrams.
This star possesses photometric characteristics which
place it exactly in the region occupied by $\lambda$ Boo stars.

Supposing that VW Ari belongs to the $\lambda$ Boo group,
one can also easily explain the remarkable difference between the metallicities
of this star and of its companion F-star with solar-like abundances
(Andrievsky et al., 1995).

\end{document}